\documentclass[english,aps,pra,reprint,noshowpacs,superscriptaddress]{revtex4-1}   
\usepackage[T1]{fontenc}	% should generally be included for better accented-word behavior
\usepackage{geometry}		% for controlling page margins
\usepackage{graphicx}
\usepackage[above,below]{placeins}	% allows use of \FloatBar­rier command to force section barriers
\usepackage{times}
\usepackage{hyperref}  
\hypersetup{colorlinks=true,urlcolor=blue,citecolor=blue,linkcolor=blue} 
\usepackage{enumitem}
\usepackage{color}
\definecolor{mygray}{gray}{0.6}

\begin{document}

\preprint{APS/123-QED}

\title{Denoting and Comparing Leadership Attributes and Behaviors in Group Work}

\author{Kristina Griswold}
\affiliation{College of Agriculture and Natural Resources, Michigan State University, East Lansing, MI 48824}
\author{Daryl McPadden}
\affiliation{Department of Physics and Astronomy, Michigan State University, East Lansing, MI 48824}
\author{Marcos D. Caballero}
\affiliation{Department of Physics and Astronomy, Michigan State University, East Lansing, MI 48824}
\affiliation{CREATE for STEM Institute, Michigan State University, East Lansing, MI 48824}
\affiliation{Department of Physics \& Center for Computing in Science Education, University of Oslo, N-0316 Oslo, Norway}
\author{Paul Irving}
\affiliation{Department of Physics and Astronomy, Michigan State University, East Lansing, MI 48824}

\begin{abstract}
\verb+Abstract.+ Projects and Practices in Physics (P$^3$) is an introductory physics class at Michigan State University that replaces lectures with a problem based learning environment. To promote the development of group based practices, students all receive group and individual feedback at the end of each week. The groups are comprised of four students, one of which often takes on the role of being the group's ``leader.'' Developing leadership based skills is a specific learning goal of the P$^3$  learning environment and the goal of this research is to examine what leadership-specific actions/traits students in P$^3$ demonstrate while working in their group. The initial phase of this study examined multiple pieces of literature to identify possible characteristics and behaviors that may present themselves in potential leaders -- creating a codebook. This phase of the study applies the codebook to in-class data to compare two tutor-labeled leaders and their leadership styles.

\end{abstract}

\maketitle

%\tableofcontents

\section{Introduction}

%Throughout history, education and all that it encompasses has been, and still is, a complex issue. With that complexity also comes numerous chances for change and improvements, or in a simpler sense; reforms. While large focuses of the past have been funding and equal right to education (both of which are still very relevant today), there has come a newer push in education, which is the STEM field. While the push for STEM fields and classes was a reform within education, there are now reforms focused on student learning within STEM, many of which are also gaining popularity in other fields/subjects. These reforms are so valued, that APS has awarded over 20 post-secondary institutions for Improving Undergraduate Physics Education \cite{APS2018AwardAwardees}. 

Projects and Practices in Physics (P$^3$), a group focused introductory college physics course, is restructuring the classroom of a typical college physics course. Divergent from the traditional lecture format, the reforms in P$^3$ center around group work and improving students' teamwork abilities. In this research, we focused on one of those abilities: leadership. It is essentially inevitable that when a group of people come together as a team, a leader emerges. There are many paths for that leader to come to be (e.g., having been appointed, self-declared, having previously held the position), but no matter which path they have taken, they have only completed half of their leadership journey. Once the leader has assumed their position, they must also work to retain it using leadership-specific actions and traits. For our purposes, the latter part of the leadership journey is especially important and is telling in how a student is performing as a leader. The actions and traits of the leader within a learning group can dramatically influence the learning of the other group members.

In this paper, we will present a framework that allows for the preliminary examination of leadership from both a positive perspective for the P$^3$ learning environment and a negative one. By positive and negative, we are focused on positive leadership influences and interactions that students can have on another student's learning in the context of P$^3$. In addition, we will compare two students portraying desired and undesired leadership in a group environment (as defined by the P$^3$ learning goals).

\section{Literature Review}
Before data analysis could begin, we needed to establish a codebook. To create our codebook, we reviewed literature that looked at the attributes, actions, ascensions, and classifications of leaders \cite{Ezzedeen2005LeadershipOrganizations,Franz,Jung}. After am extensive literature review we refined our codebook to focus on leadership methods and perspectives, leader-follower relationships, categorization, and leader identity.

The initial phase in developing our codebook was to record all of the characteristics and actions as described in leadership in group-based literature. This included information on leader-follower interactions, common characteristics of a leader, and types of leadership. For the next stage, we needed to translate the literature from its context to ours. Largely this meant shifting the context from a manager and employees situation to our situation of a student leader and other student group members.

From there, the literature was grouped using similarities in focus. That is, leadership traits were grouped together, leader-follower relations were grouped together, and so on. This was to prevent the creation of duplicate codes as well as to ensure that there were not any gaps in our literature findings. From that point, we created codes from the literature by defining them in our context, and designated what were desired and undesired ways of carrying out the codes based off of the classroom goals set by $P^3$. After reviewing our data and through the initial testing of the framework, we discovered two emergent codes that were not covered by the literature. For these codes, we also designated undesired and desired ways for them to be carried out in the P$^3$ context, just as we did for literature based codes. These can be seen in Table \ref{tab:codes} below, and are noted as emergent in their source.

\begin{table*}[t]
\caption{The following table is codebook, representing a list of all codes present in our data, their descriptions according to the context of P$^3$, and their source. The emergent codes are bold with a source of Em. The codes that show negative leadership attributes are in gray with an asterisk. Each code has appeared in Natasha's data, Logan's data, or both. Codes referenced here can also be seen in the graphs of Figure \ref{AllCodes} in the results section.\label{lm}}
\renewcommand{\arraystretch}{1.25}
\begin{tabular}{ p{1.25 in} p{2 in} p{0.3 in} | p{0.75 in} p{2 in} p{0.3 in}}
\hline
\rule{0pt}{4ex}Code  & Description and Source  &  & Code  & Description and Source &   \vspace{1mm}\\\hline
\rule{0pt}{4ex}Bought Into Group & Comments about enjoying the group, explaining to group members what they missed & \cite{Hogg2002Leader-MemberIdentity} & Contextual Buy-In To P$^3$  & Positive statements about P$^3$, Following and/or promoting structures and procedures set by P$^3$  & \cite{Hogg2002Leader-MemberIdentity}\\

Coolness Under Pressure  & Positive comments in time constraints, reassuring group members  & \cite{Chemers2003LeadershipPerspectives} & \textbf{Creating A Plan}  & \textbf{Suggesting the need for a plan, outlining the steps of a plan}  & \textbf{Em}\\

Developing Productive Relationships Within Group  & Organizing communication outside of class, asking for explanations between other group members, mediating interactions, validating group members & \cite{Chemers2003LeadershipPerspectives} & Encouraging Involvement  & Asking for individual(s) to participate, passing out materials, entertaining opposing viewpoints  & \cite{Chemers2003LeadershipPerspectives}\\

\textbf{Ensuring Shared Understanding}  & \textbf{Checking if a person(s) understands, clarifying for understanding}  & \textbf{Em} & Explaining Decision  & Explaining reasoning while suggesting a decision or after a decision has been made  & \cite{Chemers2003LeadershipPerspectives}\\

Goal Setting  & Proposing that the group needs a goal, goal is made and/or reiterated & \cite{Lord2003IdentitySchema} & Informed/ Prepared  & Referencing reading notes and homework, referencing own notes  & \cite{Lord2003IdentitySchema}\\

Insight  & Making connections between aspects of the problem  & \cite{Chemers2003LeadershipPerspectives} & Motivate Group Members  & Inspiring a change in behavior from motivational comments  & \cite{Chemers2003LeadershipPerspectives}\\

\textcolor{mygray}{Not Bought Into Group*}  & \textcolor{mygray}{Statements of disliking group work or group members, working against and/or apart from the group}  & \cite{Hogg2002Leader-MemberIdentity} & \textcolor{mygray}{Not Contextual Buy-In To P$^3$*}  & \textcolor{mygray}{Negative statements about the class, working against promoted structures and class procedures}  & \cite{Hogg2002Leader-MemberIdentity}\\

\textcolor{mygray}{Not Patient*} & \textcolor{mygray}{Talking over or interrupting group members, not explaining, pushing for a solution as quickly as possible} & \cite{Chemers2003LeadershipPerspectives,Lord2003IdentitySchema} & Prior Experience  & Making statements about previous knowledge, utilizing previous knowledge  & \cite{Chemers2003LeadershipPerspectives,Platow2003LeadershipProcesses}\\

Promoting Group Organization  & Asking for a written record of conversation both in class or by sharing materials outside of class  & \cite{Lord2003IdentitySchema} & Responsible For Group Actions  & Admitting, accepting, or reflecting on mistakes, not blaming others & \cite{Lord2003IdentitySchema} \\

Sought Out For Confirmation  & Group members repeatedly seeking approval from the same group member  & \cite{Chemers2003LeadershipPerspectives} & Task Oriented  & Suggesting to work solely on the task at hand, working with a narrow scope  & \cite{Chemers2003LeadershipPerspectives}\\

Tutor Validation Of A Leader - Tutor Initiated  & In class or via weekly feedback validation of prior experience or ideas   & \cite{Chemers2003LeadershipPerspectives} & \textcolor{mygray}{Uninformed/ Unprepared*}  & \textcolor{mygray}{Statements about not reading notes or doing homework, not knowing where to find class resources}  & \cite{Lord2003IdentitySchema}\\

\textcolor{mygray}{Unproductive Behavior*}  & \textcolor{mygray}{Saying rude comments or being demeaning to group members}  & \cite{Chemers2003LeadershipPerspectives}\\
\hline
\end{tabular}\label{tab:codes}
\end{table*}

\section{Context}
P$^3$ employs a problem-based learning (PBL) format, emboldening the students to actively engage with the material and in turn, their group members. The goals for this course focus on developing many scientific practices, which in conjunction with the format of the class, also emphasizes an improvement in group work skills. In the semester our data was collected, 40 students were enrolled in the course, which were then sorted to be ten groups of four students. Those same four students work together for the first third of the semester, until their first exam. Groups are also assigned a tutor from one of the following: a faculty member, graduate student, or undergraduate learning assistant (ULA). The tutor's role is to provide guidance during the class, give weekly written feedback about performance both as a whole and individually, as well as to grade their students' performance in class. Class is held twice per week, each session being nearly two hours. Both days, the students are equipped with white boards, markers, computers with internet access, their notes, and their tutor. On the first day of the week, students solve an analytical problem. They are given some parameters, are left with several unknowns to find. Students work their way through the problem by both using math and/or physics notes and knowledge as well as proposing theoretical experiments to find values through their tutors. The second day of the week typically builds off of the first. Sometimes it is a day focused on solving a physics problem computationally using a minimally working program. The students have to create a model by writing and repairing Python code to match what they have found during the previous class period. Because some students may come in with prior knowledge in physics and/or coding, we have collected our data from both days of the week for the purposes of this research. For a more in-depth discussion of the contextual figures of P$^3$, please refer to \citet*{0143-0807-38-5-055701}.

\section{Methods}

In order to be able to apply our codebook to the data, we needed to divide up our codes into three possible variants; bid, response, and action. We defined each as follows:

\begin{enumerate}[noitemsep,nolistsep]
\item Bid: An attempt at establishing oneself as a possible leader. 
\item Response: An external marker that demonstrates whether one's bid is acknowledged or dismissed by the group. This can also be an external validation of leadership from an authority figure.
\item Action: Demonstrating a leadership-based behavior within the classroom.
\end{enumerate}

Each code occurrence is assigned one variant (bid/response/action) in order to specify which level of leadership the student is portraying. It should be noted that not all codes have all three variants. Some codes only have one of those three as possible occurrences, while others have all three variants as possible occurrences. For example, the code Prior Experience has all three possible variants. {\it Prior Experience - Bid} is defined as trying to become a credible and legitimate source of influence. {\it Prior Experience - Response} is when another group member states/indicates that highlights another student's previous knowledge. {\it Prior Experience - Action} occurs when the student is clearly demonstrating a use of their previous knowledge. {\it Responsible For The Group's Actions} on the other hand \textit{only} has an action, defined as admitting mistakes, accepting and reflecting on mistakes, not blaming others.

Ideally, this framework would allow us to code for the whole interaction between the leader and at least one of their group members. While that is the ideal circumstance, we recognize that our research is on humans and not a perfect scenario. Because of this, not every interaction will utilize this whole sequence. Many of the interactions showcase only one variant: bid, validation, or action. Nevertheless, there is a significant difference between the three variants that warrants distinction. 

An example of a whole bid-response-action interaction could exist as the following:

\begin{quote}
\textbf{Student 1:} I have never done any coding before, so I don't know how to do today's project.\\
\textbf{Student 2:} Oh, I have taken a computer science course. I've got some experience with C++, but not Python. I still can probably understand the code and how it works though.\\
\textbf{Student 1:} Good, at least one of us knows something about coding.\\
\textbf{Student 3:} Yeah, because I haven't coded before either.\\
\textbf{Student 2:} That's okay. I think I can code this part with the while-loop (begins to code). 
\end{quote}

Student 2's first remark would be an example of the bid because they are stating their prior experience; a possible attempt to establish themselves as a leader. The next remarks from Students 1 and 3 would be examples of responses as they show that Student 2's comment was heard; the bid was acknowledged. Then Student 2's start of coding would be an example of an action because they begin to demonstrate their knowledge by editing the code.

In relation to Fig.~\ref{AllCodes} in Sec.~\ref{sec:results}, only bid and action are counted towards the data as the responses do not directly correlate to a student's attempt at leadership. There are a few exceptions to this, which are the codes {\it Prior Experience}, {\it Tutor Validation Of A Leader - Tutor Initiated}, and {\it Informed/Prepared For Class}. These codes are being counted towards the total because they show direct evidence that the student possesses a leadership trait. We will be using data from the response variants in the future to look at how many times a student has a bid compared to how many times the response indicates that bid was validated. 

As for our dataset, we are looking at in-class video data of two separate groups, which were drawn from the same class. Both groups also had the same tutor, and female, male breakdown (2 x 2). Within those groups, we are looking specifically at two students (one per group), whose pseudonyms are Natasha and Logan. These two students have been chosen because their tutor had indicated in their weekly feedback to them and/or in interviews with that tutor that they were the leaders of their respective groups. Because they have been identified as leaders by the tutor, we are interested in applying the framework to them to understand what characteristics and actions they took as leaders. As mentioned previously, one portrays a positive leadership style for their group environment (Natasha), while the other portrays a negative one (Logan). This makes their comparison more drastic and dynamic than that of leaders with similar leadership styles.

\section{Results and Discussion}\label{sec:results}

\begin{figure*}
\begin{center}
\includegraphics[scale=0.45]{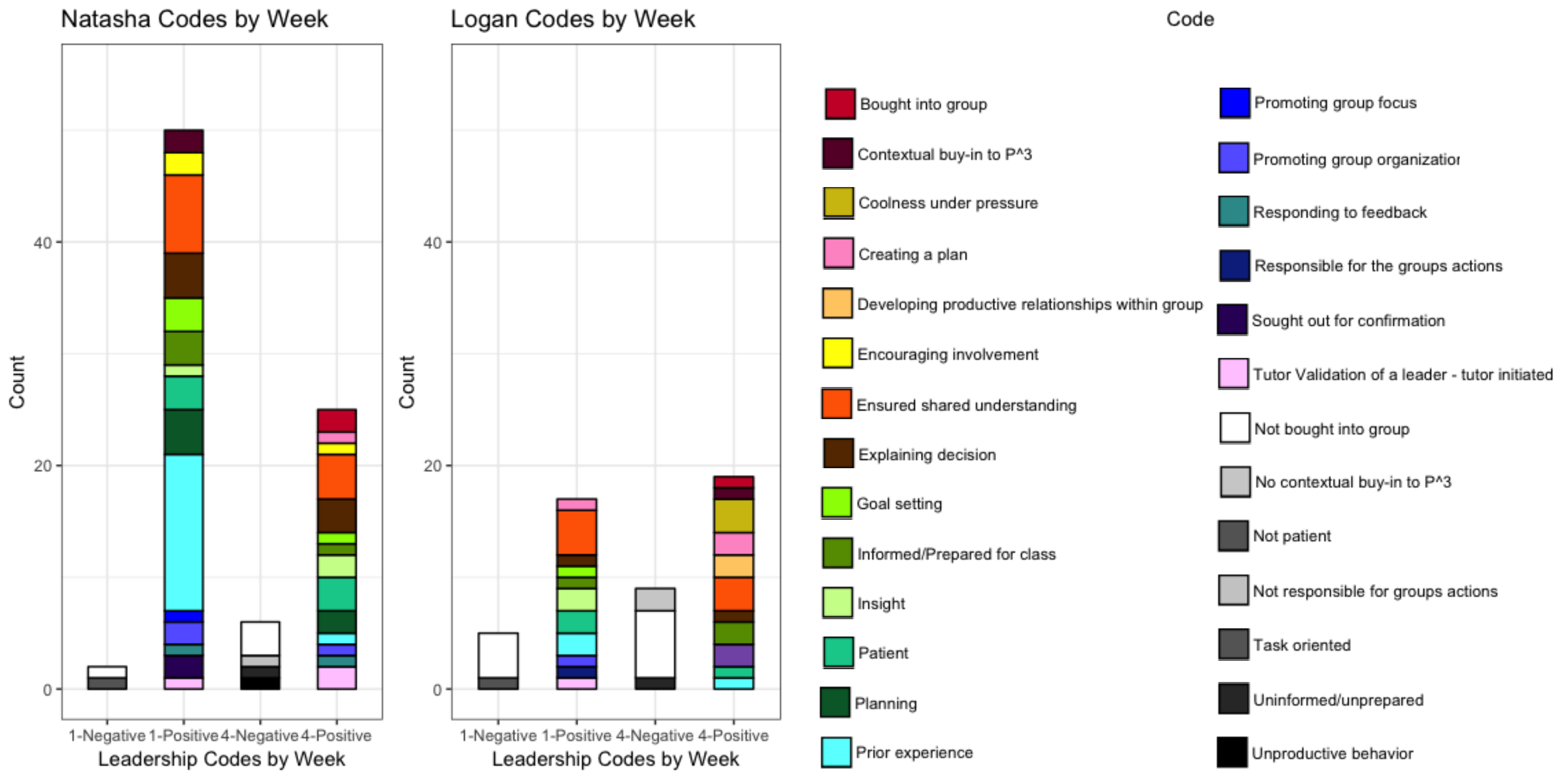}
\end{center}
\caption{This figure shows the leadership codes for Natasha and Logan during the first and fourth weeks of $P^3$. \label{AllCodes}}
\end{figure*}

Fig.~\ref{AllCodes} shows all of the codes from both Week 1 and Week 4 for Natasha and Logan. The graphs are formatted so that the first column of each week is negative codes, and the second is positive. Each color represents a specific code, which is listed next to the plots in Fig.~\ref{AllCodes}.

Looking at Fig.~\ref{AllCodes}, it is evident that Natasha progresses from a group member who states and uses her prior experience as a large portion of her leadership codes to one who utilizes that information to help her group members. In Week 1, we can see that she has roughly 10 of her 50 code occurrences appearing in {\it Prior Experience}, with the next largest sections being {\it Ensuring Shared Understanding}, {\it Explaining Decision}, and {\it Informed/Prepared for Class}. Moving to Week 4, Natasha all but eliminates her use of {\it Prior Experience} in exchange for adding in codes like {\it Creating a Plan} and {\it Bought into Group}, showing that she is working on facilitating rather than leading her group by doing much of the work. Though she does show an increase of negative codes from Week 1 to Week 4, her ratio is a maximum of 1 negative code to 4 positive codes. In both Week 1 and Week 4, Natasha demonstrates 15 positive leadership codes, showing a wide variety of codes each week.

For Logan (shown in Fig.~\ref{AllCodes}), there is a slight increase of both positive and negative code occurrences from Week 1 to Week 4, with the maximum ratio being 1 negative code to 2 positive codes. In both Week 1 and Week 4, Logan demonstrates 11 positive leadership codes, showing a slight variety of codes each week. His number of positive codes remains relatively constant, showing a lack of adaptability. Comparing Logan's data to Natasha's, it is evident that Natasha is much more active as a leader than Logan is. With that activity, also comes fewer negative codes both proportionally as well as in pure numbers.

Using these initial findings, we aim to apply the framework to the other students in Natasha's and Logan's groups to evaluate whether the framework clearly illustrates differences in group member leadership. In addition, we hope to expand the study to see if and how the weekly feedback from the tutors impacts a leader's behavior. Ideally, that feedback would move positive leaders to a facilitation role (much like the progression of Natasha's data in this study) and move negative leaders to a positive leader position. This continuation of the study would allow us to test our belief that the feedback is beneficial to students' teamwork skills in the class. It is important to note that our research does not directly take into account challenges that arise with leadership such as race and gender. While those are both certainly factors that have to potential to drastically change the outcome of who is a leader and who is not, that is beyond the scope of the study. But, both will be important to examine in the future. The findings of this study are congruent with the statements made by Natasha and Logan's tutor, as the tutor depicted Natasha as a positive leader and Logan as a negative one. In turn, this demonstrates that perceived leaders can retain power in their groups despite portraying remarkably different leadership styles in the same environment. 

%If true, we would demonstrate that tutor feedback is an important step to incorporate in the physics classroom and beyond.

%\bibliography{ModelingNetworkRepresentation}
\bibliography{Mendeley}

\begin{thebibliography}{8}%
\makeatletter
\providecommand \@ifxundefined [1]{%
 \@ifx{#1\undefined}
}%
\providecommand \@ifnum [1]{%
 \ifnum #1\expandafter \@firstoftwo
 \else \expandafter \@secondoftwo
 \fi
}%
\providecommand \@ifx [1]{%
 \ifx #1\expandafter \@firstoftwo
 \else \expandafter \@secondoftwo
 \fi
}%
\providecommand \natexlab [1]{#1}%
\providecommand \enquote  [1]{``#1''}%
\providecommand \bibnamefont  [1]{#1}%
\providecommand \bibfnamefont [1]{#1}%
\providecommand \citenamefont [1]{#1}%
\providecommand \href@noop [0]{\@secondoftwo}%
\providecommand \href [0]{\begingroup \@sanitize@url \@href}%
\providecommand \@href[1]{\@@startlink{#1}\@@href}%
\providecommand \@@href[1]{\endgroup#1\@@endlink}%
\providecommand \@sanitize@url [0]{\catcode `\\12\catcode `\$12\catcode
  `\&12\catcode `\#12\catcode `\^12\catcode `\_12\catcode `\%12\relax}%
\providecommand \@@startlink[1]{}%
\providecommand \@@endlink[0]{}%
\providecommand \url  [0]{\begingroup\@sanitize@url \@url }%
\providecommand \@url [1]{\endgroup\@href {#1}{\urlprefix }}%
\providecommand \urlprefix  [0]{URL }%
\providecommand \Eprint [0]{\href }%
\providecommand \doibase [0]{http://dx.doi.org/}%
\providecommand \selectlanguage [0]{\@gobble}%
\providecommand \bibinfo  [0]{\@secondoftwo}%
\providecommand \bibfield  [0]{\@secondoftwo}%
\providecommand \translation [1]{[#1]}%
\providecommand \BibitemOpen [0]{}%
\providecommand \bibitemStop [0]{}%
\providecommand \bibitemNoStop [0]{.\EOS\space}%
\providecommand \EOS [0]{\spacefactor3000\relax}%
\providecommand \BibitemShut  [1]{\csname bibitem#1\endcsname}%
\let\auto@bib@innerbib\@empty
%</preamble>
\bibitem [{\citenamefont
  {Ezzedeen}(2005)}]{Ezzedeen2005LeadershipOrganizations}%
  \BibitemOpen
  \bibfield  {author} {\bibinfo {author} {\bibfnamefont {S.~R.}\ \bibnamefont
  {Ezzedeen}},\ }\href {\doibase
  http://dx.doi.org/10.1111/j.1744-6570.2005.20050209{\_}4.x} {\bibfield
  {journal} {\bibinfo  {journal} {Personnel Psychology}\ } (\bibinfo {year}
  {2005}),\
  http://dx.doi.org/10.1111/j.1744-6570.2005.20050209{\_}4.x}\BibitemShut
  {NoStop}%
\bibitem [{\citenamefont {Franz}\ and\ \citenamefont {James
  R.~Larson}(2002)}]{Franz}%
  \BibitemOpen
  \bibfield  {author} {\bibinfo {author} {\bibfnamefont {T.~M.}\ \bibnamefont
  {Franz}}\ and\ \bibinfo {author} {\bibfnamefont {J.}~\bibnamefont {James
  R.~Larson}},\ }\href {\doibase 10.1177/104649640203300401} {\bibfield
  {journal} {\bibinfo  {journal} {Small Group Research}\ }\textbf {\bibinfo
  {volume} {33}},\ \bibinfo {pages} {383} (\bibinfo {year} {2002})}\BibitemShut
  {NoStop}%
\bibitem [{\citenamefont {Jung}\ and\ \citenamefont {Sosik}(2002)}]{Jung}%
  \BibitemOpen
  \bibfield  {author} {\bibinfo {author} {\bibfnamefont {D.~I.}\ \bibnamefont
  {Jung}}\ and\ \bibinfo {author} {\bibfnamefont {J.~J.}\ \bibnamefont
  {Sosik}},\ }\href {\doibase 10.1177/10496402033003002} {\bibfield  {journal}
  {\bibinfo  {journal} {Small Group Research}\ }\textbf {\bibinfo {volume}
  {33}},\ \bibinfo {pages} {313} (\bibinfo {year} {2002})}\BibitemShut
  {NoStop}%
\bibitem [{\citenamefont {Hogg}\ \emph {et~al.}(2002)\citenamefont {Hogg},
  \citenamefont {Martin},\ and\ \citenamefont
  {Weeden}}]{Hogg2002Leader-MemberIdentity}%
  \BibitemOpen
  \bibfield  {author} {\bibinfo {author} {\bibfnamefont {M.~A.}\ \bibnamefont
  {Hogg}}, \bibinfo {author} {\bibfnamefont {R.}~\bibnamefont {Martin}}, \ and\
  \bibinfo {author} {\bibfnamefont {K.}~\bibnamefont {Weeden}},\ }in\
  \href@noop {} {\emph {\bibinfo {booktitle} {Leadership and Power: Identity
  Processes in Groups and Organizations}}},\ \bibinfo {editor} {edited by\
  \bibinfo {editor} {\bibfnamefont {D.}~\bibnamefont {van Kippenberg}}\ and\
  \bibinfo {editor} {\bibfnamefont {M.~A.}\ \bibnamefont {Hogg}}}\ (\bibinfo
  {publisher} {SAGE Publications},\ \bibinfo {address} {London (ENG), Thousand
  Oaks (CA), New Delhi (IND)},\ \bibinfo {year} {2002})\ \bibinfo {edition}
  {1st}\ ed.,\ Chap.~\bibinfo {chapter} {3}, pp.\ \bibinfo {pages}
  {18--33}\BibitemShut {NoStop}%
\bibitem [{\citenamefont {Chemers}(2003)}]{Chemers2003LeadershipPerspectives}%
  \BibitemOpen
  \bibfield  {author} {\bibinfo {author} {\bibfnamefont {M.~M.}\ \bibnamefont
  {Chemers}},\ }in\ \href@noop {} {\emph {\bibinfo {booktitle} {Leadership and
  Power: Identity Processes in Groups and Organizations}}},\ \bibinfo {editor}
  {edited by\ \bibinfo {editor} {\bibfnamefont {D.}~\bibnamefont {van
  Kippenberg}}\ and\ \bibinfo {editor} {\bibfnamefont {M.~A.}\ \bibnamefont
  {Hogg}}}\ (\bibinfo  {publisher} {SAGE Publications},\ \bibinfo {address}
  {London (ENG), Thousand Oaks (CA), New Delhi (IND)},\ \bibinfo {year}
  {2003})\ \bibinfo {edition} {1st}\ ed.,\ Chap.~\bibinfo {chapter} {2}, pp.\
  \bibinfo {pages} {5--17}\BibitemShut {NoStop}%
\bibitem [{\citenamefont {Lord}\ and\ \citenamefont
  {Hall}(2003)}]{Lord2003IdentitySchema}%
  \BibitemOpen
  \bibfield  {author} {\bibinfo {author} {\bibfnamefont {R.}~\bibnamefont
  {Lord}}\ and\ \bibinfo {author} {\bibfnamefont {R.}~\bibnamefont {Hall}},\
  }in\ \href@noop {} {\emph {\bibinfo {booktitle} {Leadership and Power:
  Identity Processes in Groups and Organizations}}},\ \bibinfo {editor} {edited
  by\ \bibinfo {editor} {\bibfnamefont {D.}~\bibnamefont {van Kippenberg}}\
  and\ \bibinfo {editor} {\bibfnamefont {M.~A.}\ \bibnamefont {Hogg}}}\
  (\bibinfo  {publisher} {SAGE Publications},\ \bibinfo {address} {London
  (ENG), Thousand Oaks (CA), New Delhi (IND)},\ \bibinfo {year} {2003})\
  \bibinfo {edition} {1st}\ ed.,\ Chap.~\bibinfo {chapter} {5}, pp.\ \bibinfo
  {pages} {48--64}\BibitemShut {NoStop}%
\bibitem [{\citenamefont {Platow}\ \emph {et~al.}(2003)\citenamefont {Platow},
  \citenamefont {Haslam}, \citenamefont {Foddy},\ and\ \citenamefont
  {Grace}}]{Platow2003LeadershipProcesses}%
  \BibitemOpen
  \bibfield  {author} {\bibinfo {author} {\bibfnamefont {M.~J.}\ \bibnamefont
  {Platow}}, \bibinfo {author} {\bibfnamefont {A.}~\bibnamefont {Haslam}},
  \bibinfo {author} {\bibfnamefont {M.}~\bibnamefont {Foddy}}, \ and\ \bibinfo
  {author} {\bibfnamefont {D.~M.}\ \bibnamefont {Grace}},\ }in\ \href@noop {}
  {\emph {\bibinfo {booktitle} {Leadership and Power: Identity Processes in
  Groups and Organizations}}},\ \bibinfo {editor} {edited by\ \bibinfo {editor}
  {\bibfnamefont {D.}~\bibnamefont {van Kippenberg}}\ and\ \bibinfo {editor}
  {\bibfnamefont {M.~A.}\ \bibnamefont {Hogg}}}\ (\bibinfo  {publisher} {SAGE
  Publications},\ \bibinfo {address} {London (ENG), Thousand Oaks (CA), New
  Delhi (IND)},\ \bibinfo {year} {2003})\ \bibinfo {edition} {1st}\ ed.,\
  Chap.~\bibinfo {chapter} {4}, pp.\ \bibinfo {pages} {34--37}\BibitemShut
  {NoStop}%
\bibitem [{\citenamefont {Irving}\ \emph {et~al.}(2017)\citenamefont {Irving},
  \citenamefont {Obsniuk},\ and\ \citenamefont
  {Caballero}}]{0143-0807-38-5-055701}%
  \BibitemOpen
  \bibfield  {author} {\bibinfo {author} {\bibfnamefont {P.~W.}\ \bibnamefont
  {Irving}}, \bibinfo {author} {\bibfnamefont {M.~J.}\ \bibnamefont {Obsniuk}},
  \ and\ \bibinfo {author} {\bibfnamefont {M.~D.}\ \bibnamefont {Caballero}},\
  }\href {http://stacks.iop.org/0143-0807/38/i=5/a=055701} {\bibfield
  {journal} {\bibinfo  {journal} {European Journal of Physics}\ }\textbf
  {\bibinfo {volume} {38}},\ \bibinfo {pages} {55701} (\bibinfo {year}
  {2017})}\BibitemShut {NoStop}%
\end{thebibliography}%

\end{document}